\documentstyle[twoside,fleqn,npb
]{article}
\newcommand{\bea}{\begin{eqnarray}}
\newcommand{\bq}{\begin{equation}}
\newcommand{\eea}{\end{eqnarray}}
\newcommand{\eq}{\end{equation}}

\newcommand\xx{\tilde{x}}

\newcommand{\AmS}{{\protect\the\textfont2
A\kern-.1667em\lower.5ex\hbox{M}\kern-.125emS}}

\title{Quantum Field Theoretic Treatment of the Non--Forward Compton
       Amplitude in the Generalized Bjorken Region
}
\author{Johannes Bl\"umlein
        \address{Deutsches Elektronen--Synchrotron,        DESY,
     Platanenallee 6, D--15735 Zeuthen, Germany}%
     \thanks{
                   DESY 00-045,~~Contribution to the Proceedings
          of `Loops and Legs in Quantum Field Theory', April 2000,
          Bastei, Germany, Nucl. Phys.  B (Proc. Suppl.), (2000)
          to appear}
        Bodo Geyer, Markus Lazar\address{
        Institut f\"ur Theoretische Physik, 
        Universit\"at Leipzig,  
        Augustusplatz 10, D 04109 Leipzig, Germany}
        and
        Dieter Robaschik\address{Institut f\"ur Theoretische Physik, 
        Karl--Franzens--Universit\"at Graz, 
        Universit\"atsplatz  5, A--8010 Graz, Austria}%
}
\begin{document}

\begin{abstract}
\noindent
A quantum field theoretic treatment of the leading light--cone part of
the virtual Compton amplitude is presented. The twist--decomposition
of the operators is performed by a group--theoretic procedure respecting
the Lorentz group $O(3,1)$. The twist--2 contributions to the Compton
amplitude are calculated and it is shown that the electromagnetic current
is conserved for these terms. Relations between the amplitude functions
associated to the symmetric and asymmetric part of the Compton amplitude
are derived. These relations generalize the Callan--Gross and
Wandzura--Wilczek relations of forward scattering for the non--forward
Compton amplitude.
\end{abstract}

\maketitle


\section{INTRODUCTION}
\renewcommand{\theequation}{\thesection.\arabic{equation}}
\setcounter{equation}{0}
\label{sec-1}
\vspace{1mm}
\noindent
Compton scattering of a virtual photon off a hadron
$\gamma^*_1 + p_1 \rightarrow {\gamma'}_2^{*} + p_2 $
is an important processes in QCD:
theoretically it can be treated in detail and
experimentally it can be tested for a large variety of processes.
In lowest approximation in the electromagnetic coupling 
it is described by
\bea
\label{COMP}
\lefteqn{
T_{\mu\nu}(p_+,p_-,q)
= i \int d^4x \,e^{iqx}\,\times}\\
&\qquad
\langle p_2, S_2\,|T (J_{\mu}(x/2) J_{\nu}(-x/2))|\,p_1, S_1\rangle\ ,
\nonumber
\eea
where
\begin{eqnarray}
p_\pm = p_2 \pm p_1, \quad
q = \hbox{\large $\frac{1}{2}$} \left(q_1 + q_2\right),
\nonumber
\end{eqnarray}
with $q_1~(q_2)$ and $p_1~(p_2)$ being the four--momenta of the
incoming~(outgoing) photon and hadron, respectively, and $S_{1}, S_{2}$
being the spins of the initial-- and final--state hadron, where
$p_1 + q_1 = p_2 + q_2$.
The generalized Bjorken region is the asymptotic domain being defined by
\begin{eqnarray}
\label{gBr1}
\nu =  qp_+ \longrightarrow \infty, \qquad
- q^2 \longrightarrow \infty~,
\nonumber
\end{eqnarray}
where the two scaling variables
\begin{eqnarray}
\label{gBr2}
\xi  = - \frac{q^2 }{qp_+}, \qquad
\eta = \frac{qp_-}{qp_+} = \frac{q_1^2 - q_2^2}{2\nu}
\end{eqnarray}
are fixed. Of special experimental importance are the cases of deep
inelastic scattering (DIS) described by the absorptive part of the 
forward Compton amplitude, $\eta = 0$, and the deeply virtual Compton 
scattering (DVCS)
with one real outgoing photon $ q_2^2 =0 $ corresponding to
$\xi = -\eta$. 
In the generalized Bjorken region the amplitude (\ref{COMP}) is
dominated by the light--cone singularities which allows to apply
the (non--local) operator product expansion of $T(J_\mu(x/2)J_\nu(-x/2))$.

Our aim is a detailed quantum field theoretic investigation of that 
approach, 
cf. Refs. \cite{BGR,BR}. In contrast to earlier considerations 
\cite{LEIP,ADI} 
we take into account the explicit twist decomposition of the {\em 
non--local
vector operators}, and their matrix elements, which thereby occur. The 
twist 
decomposition in the case of the quark--antiquark operators has been 
treated 
in \cite{GLR}. An extension to the gluon operators and more general 
multiparticle
operators is given in \cite{GL}. The relations between vector and scalar 
operators of twist 2  allows to express the final results with the help 
of
matrix elements of the scalar operators only. In addition, this procedure 
leads
to the derivation of new relations~\cite{BR} on the amplitude level which 
correspond in the case of forward scattering to the Callan-Gross and 
Wandzura-Wilczek relations. Furthermore, the electromagnetic current 
conservation
may be shown to hold on the level of the twist 2 contributions.
\section{LIGHT--CONE--EXPANSION}
\renewcommand{\theequation}{\thesection.\arabic{equation}}
\setcounter{equation}{0}
\label{sec-2}

The Compton amplitude for the case of non--forward scattering,
Eq.~(\ref{COMP}), in the generalized Bjorken region is dominated by the 
light--cone singularities. Therefore, the $T$--product of the
electromagentic currents will be approximated by its non--local
light--cone expansion~\cite{NLC} -- a summed--up
form of the local light--cone expansion which allows for a quite compact 
representation of the resulting expressions. 

Let us present a series of intuitive arguments leading to that
approximation. We start from the renormalized time-ordered operator
product ($S$ being the renormalized $S$--matrix):
\begin{eqnarray}
\widehat{T}_{\mu\nu}(x)  =
i RT \left[J_\mu\left({x}/{2}\right)
J_\nu\left(-{x}/{2}\right) S \right]\ .
\nonumber
\end{eqnarray}
At first we consider this expression in the Born approximation
\begin{eqnarray}
\lefteqn{
\widehat{T}_{\mu\nu}(x)  =
    -e^2 \frac{ x^\lambda}{2 \pi^2 (x^2-i\epsilon)^2}\,\times
}\nonumber\\
& \left[
\overline{\psi}
\big(\hbox{\large$\frac{ x}{2}$}\big)
\gamma_\mu \gamma_\lambda \gamma_\nu \psi
\big(\hbox{\large$\frac{-x}{2}$}\big)
- \overline{\psi}
\big(\hbox{\large$\frac{-x}{2}$}\big)
\gamma_\nu \gamma_\lambda \gamma_\mu \psi
\big(\hbox{\large$\frac{x}{2}$}\big)
\right]~.
\nonumber
\end{eqnarray}
Here $e$ denotes the charge of the fermion
field $\psi$. We dropped the flavor indices in the expressions considered.
Reordering in the standard way the Dirac--structure we obtain
\begin{eqnarray}
\label{Tmunug}
\lefteqn{
 \widehat{T}_{\mu\nu}(x)  =
 - e^2 \frac{x^\lambda}{i \pi^2 (x^2-i\epsilon)^2}\,\times
}\\
&\quad \left[S_{\alpha \mu\lambda \nu} 
 O^\alpha \big(\hbox{\large$\frac{x}{2}$},\hbox{\large$\frac{-x}{2}$}\big)
-i \varepsilon_{\alpha\mu\lambda\nu } 
O_5^\alpha \big(\hbox{\large$\frac{x}{2}$},\hbox{\large$\frac{-x}{2}$}\big)
\right],
\nonumber
\end{eqnarray}
where
\begin{eqnarray}
S_{\alpha \mu\lambda \nu} = g_{\alpha\mu}g_{\lambda \nu}
                          + g_{\lambda\mu}g_{\alpha \nu}
                          - g_{\mu\nu}g_{\lambda \alpha}\ .
\nonumber
\end{eqnarray}
The essential objects are the bilocal operators
\bea
\label{oo}
\hspace{-.75cm}
&&O^{\alpha}\big(\hbox{\large$\frac{x}{2}$},\hbox{\large$\frac{-x}{2}$}\big)
= \\
\hspace{-.75cm}
&&\quad
=\hbox{\large$\frac{i}{2}$}
\left[\overline{\psi}\big(\hbox{\large$\frac{x}{2}$}\big)
   \gamma^\alpha\psi\big(\hbox{\large$\frac{-x}{2}$}\big)
   - \overline{\psi}\big(\hbox{\large$\frac{-x}{2}$}\big)
    \gamma^\alpha\psi\big(\hbox{\large$\frac{x}{2}$}\big)\right],
\nonumber\\
\label{oo5}
\hspace{-.75cm}
&&O^{\alpha}_5
\big(\hbox{\large$\frac{x}{2}$},\hbox{\large$\frac{-x}{2}$}\big)
=\\
\hspace{-.75cm}
&&\quad
=\hbox{\large$\frac{i}{2}$}
\left[\overline{\psi}\big(\hbox{\large$\frac{x}{2}$}\big)
\gamma_5\gamma^\alpha\psi\big(\hbox{\large$\frac{-x}{2}$}\big)
+ \overline{\psi}\big(\hbox{\large$\frac{-x}{2}$}\big)
\gamma_5\gamma^\alpha\psi\big(\hbox{\large$\frac{x}{2}$}\big)\right].
\nonumber
\eea
Expression (\ref{Tmunug}) satisfies electromagnetic current conservation
in the case of free fields, i.e. at zeroth order in QCD. We are 
interested,
however, in the case of general fields $\psi$ and the twist--2
operators associated to them. We calculate these operators  at leading
order passing the following steps.

\noindent
STEP 1:  Use gauge invariant operators in place of (\ref{oo}), 
(\ref{oo5}). 
This is achieved by including the phase factor
$U(y,z) = {\cal P}\exp (ig\int_z^y A_\mu dx^\mu) $. The integration
can be performed over a straight path connecting  $y$ and $z$.\\
\noindent
STEP 2: Perform the twist decomposition of these operators according to 
\cite{GLR} and restrict to the twist--2 (axial) vector operators only. \\
\noindent
STEP 3: Take the operators as renormalized ones (at the light--cone)
to all orders of QCD:
\begin{eqnarray}
\label{oor}
\hspace{-.5cm}&&
O_{\alpha}^{\rm tw2}
\big(\hbox{\large$\frac{\xx}{2}$},\hbox{\large$\frac{-\xx}{2} $}\big)
= 
\\
\hspace{-.5cm}&&
\qquad
= \hbox{\large$\frac{i}{2}$} RT
\Big\{ 
\left[\overline{\psi}\big(\hbox{\large$\frac{\xx}{2}$}\big)
\gamma_\alpha 
U(\hbox{\large$\frac{\xx}{2}$},\hbox{\large$\frac{- \xx}{2} $}) 
\psi\big(\hbox{\large$\frac{-\xx}{2}$}\big)\right.
\nonumber\\
\hspace{-.5cm}&&
\qquad\left.
-\, \overline{\psi}\big(\hbox{\large$\frac{-\xx}{2}$}\big)
\gamma_\alpha 
U(\hbox{\large$\frac{-\xx}{2}$},\hbox{\large$\frac{\xx}{2} $})
\psi\big(\hbox{\large$\frac{x}{2}$}\big)\right]^{\rm tw2} ~S \Big\},
\nonumber\\
\label{oo5r}
\hspace{-.5cm}&&
O_{5\ \alpha}^{\rm tw2}
\big(\hbox{\large$\frac{\xx}{2}$},\hbox{\large$\frac{-\xx}{2}$}\big)
=
\\
\hspace{-.5cm}&&
\qquad
= \hbox{\large$\frac{i}{2} $} RT
\Big\{ 
\left[\overline{\psi}\big(\hbox{\large$\frac{\xx}{2}$}\big)
\gamma_5\gamma_\alpha  
U(\hbox{\large$\frac{\xx}{2}$},\hbox{\large$\frac{-\xx}{2}$}) 
\psi\big(\hbox{\large$\frac{-\xx}{2}$}\big)\right.
\nonumber\\
\hspace{-.5cm}&&
\qquad\left.
+\, \overline{\psi}\big(\hbox{\large$\frac{-\xx}{2}$}\big)
\gamma_5\gamma_\alpha 
U(-\hbox{\large$\frac{\xx}{2}$},\hbox{\large$\frac{\xx}{2} $})
\psi\big(\hbox{\large$\frac{\xx}{2}$}\big)\right]^{\rm tw2}~ S\Big\},
\nonumber
\end{eqnarray}
where
\begin{eqnarray}
\label{xtil}
\tilde x = x + {\zeta}\big[ \sqrt{(x\zeta)^2 - x^2 \zeta^2}
- (x\zeta)\big]/{\zeta^2}
\nonumber
\end{eqnarray}
and $\zeta$ is a subsidiary vector.
Therefore our final expression to be  considered in the following reads
\begin{eqnarray}
\label{Tmunur2}
\lefteqn{
 \widehat{T}^{\rm tw2}_{\mu\nu}(x)  =
 - e^2 \frac{\tilde x^\lambda}{i \pi^2 (x^2-i\epsilon)^2}\,\times
}\\
& \Big[S^\alpha_{~\mu\lambda \nu} 
O_{\alpha}^{\rm tw2}
\big(\hbox{\large$\frac{\xx}{2}$},\hbox{\large$\frac{-\xx}{2}$}\big)
-i \varepsilon^\alpha_{~\mu\lambda \nu} 
O_{5\ \alpha}^{\rm tw2}
\big(\hbox{\large$\frac{\xx}{2}$},\hbox{\large$\frac{-\xx}{2}$}\big)
\Big].
\nonumber
\end{eqnarray}
\section{TWIST DECOMPOSITION}
\renewcommand{\theequation}{\thesection.\arabic{equation}}
\setcounter{equation}{0}
\label{sec-3}
We are confronted now with the twist decomposition of 
non--local operators.
The concept of twist was originally introduced in \cite{TW} and 
successfully applied in \cite{RC}. The twist decomposition of simple
non--local operators                    was studied in \cite{BB}.
for the first time.
A unique
group theoretical procedure, based on the decomposition of related 
{\em local} tensor operators into irreducible ones with respect to the 
Lorentz group $O(3,1)$  has been introduced recently and successfully
applied to non--local tensor operators up to second rank \cite{GLR,GL}.

As an example let us consider the following (uncentered, unsymmetrized) 
quark operator
\begin{eqnarray}
O_\Gamma\left(0, \kappa x \right)
=
\left[\overline{\psi}(0) \Gamma
 U(0,\kappa x) \psi(\kappa x) \right]\ ,
\nonumber
\end{eqnarray}
where $\Gamma =\{ 1, \gamma_5; \gamma_\alpha, \gamma_\alpha \gamma_5;
\sigma_{\alpha\beta}\}$. Its expansion into local operators reads
\begin{eqnarray}
\label{oorx}
\lefteqn{
O_\Gamma\left(0, \kappa x \right)
= \sum_{n=0}^{\infty} \frac{\kappa^n}{n!} x^{\mu_1} x^{\mu_2}.... x^{\mu_n}
\,\times}\\
&\quad 
\left[\overline{\psi}(0) \Gamma 
  D_{\mu_1}(y) D_{\mu_2}(y).... D_{\mu_n}(y) \psi(y)\right]\big|_{y=0}\ ,
\nonumber
\end{eqnarray}
where $D_\mu(y)$ denotes the covariant derivative taken at $y$.
Now, the local operators whose tensor structure is determined by $\Gamma$,
e.g.,  $\alpha\beta(\mu_1 \ldots\mu_n)$, and being totally symmetric with 
respect to $\mu_1 \ldots \mu_n$ are to be decomposed into irreducible 
tensors.
These tensors should be traceless and their symmetry behavior is 
uniquely
determined by Young patterns $(m_1, m_2, \ldots m_r)$ whose $i$-th row 
has 
length $m_i$. In the (pseudo) scalar case the only allowed Young pattern 
is $(n)$, in the (axial) vector case there are two Young patterns $(n+1)$
and $(n,1)$, whereas in the antisymmetric resp. symmetric tensor case the 
Young patterns $(n+1,1)$ and $(n,1,1)$ resp.~$(n+2), (n+1,1)$ and $(n,2)$ 
appear. The complete decomposition of the local tensors into irreducible
ones besides the leading twist part contains also irreducible tensors of 
higher twist being related to the trace terms (e.g., in the symmetric 
tensor case contributions up to twist 6 occur, cf.
\cite{GL}). The next step 
to be performed
consists in resuming, according to (\ref{oorx}), the towers 
(with respect to $n$) of the local operators with the same twist to 
non--local operators of definite twist which are tensorial harmonic
functions. Let us remark that according to this definition the twist 
decomposition depends on the basic point $y=0$ where the local expansion 
is made. Finally, these non--local operators are to be projected onto the
light--cone in order to obtain the twist decomposition of the light--ray 
operators we are seeking for.

Now we list those results of \cite{GLR} which are relevant for the present
consideration:\\
\noindent
(a) Twist--2 scalar quark operators for arbitrary positions 
$\kappa_1 x$ and $ \kappa_2 x$ are given by
\begin{eqnarray}
\label{ortl}
\hspace{-.7cm}
&&O^{\rm tw2}(\kappa_1 x, \kappa_2 x )
=
O(\kappa_1 x, \kappa_2 x )
\\
\hspace{-.7cm}
&&
+\sum_{k=1}^\infty \int_0^1 \!\!\! dt 
\Big(\!\frac{1\!-\!t}{t}\!\Big)^{\!k-1} 
\frac{(-x^2)^k\Box^k}{4^k k!(k-1)!} O(\kappa_1 tx, \kappa_2 tx ) ,
\nonumber
\end{eqnarray}
with (compare Eq.~(\ref{oo}))
\begin{eqnarray}
\label{oortu}
\hspace{-.7cm}
&&
O(\kappa_1 x, \kappa_2 x)
= 
\\
\hspace{-.7cm}
&&
\qquad
=\,\hbox{\large$\frac{i}{2}$} RT
\big\{ 
\left[\overline{\psi}(\kappa_1 x)
(x \gamma) U(\kappa_1 x, \kappa_2 x) \psi(\kappa_2 x)\right.
\nonumber\\
\hspace{-.7cm}
&&
\qquad\qquad\left.
- \overline{\psi}(\kappa_2 x)
(x \gamma) U(\kappa_2 x, \kappa_1 x)
\psi(\kappa_1 x)\right] S \big\}\ .
\nonumber
\end{eqnarray}
\noindent
The operator (\ref{ortl}) being of leading twist, i.e., with all traces 
being subtracted, obeys the following important relation
\begin{eqnarray}
\label{co1}
\Box O^{\rm tw2}(\kappa_1 x,\kappa_2 x) = 0\ .
\end{eqnarray}
Obviously, on the light--cone, $x\rightarrow \xx$, both operators, 
(\ref{ortl}) and (\ref{oortu}), coincide.\\
(b) Twist--2 vector quark operators are shown to be determined through
the twist--2 scalar quark operator by \cite{GLR}
\begin{eqnarray}
\label{optw2}
O_{\alpha}^{\rm tw2} ( \kappa_1 x, \kappa_2 x)
 = \! \int_0^{1} \!\!\!d{\tau} \;
\partial_\alpha
O^{\rm tw2}(\kappa_1 \tau x, \kappa_2\tau x) ,
\end{eqnarray}
which, after projecting onto the light--cone, is crucial for the 
reduction 
of the matrix elements of the operators (\ref{oor}, \ref{oo5r}) to those 
of
the corresponding (pseudo)scalar operators. The operator (\ref{optw2})
satisfies the relations
\begin{eqnarray}
\label{co3}
&&\hspace{-.7cm}
\Box O^{\rm tw2}_\alpha(\kappa_1 x,\kappa_2 x) = 0\ ,
\\
\label{co2}
&&\hspace{-.7cm}
\partial^\alpha O^{\rm tw2}_\alpha(\kappa_1 x,\kappa_2 x) = 0\ .
\end{eqnarray}
The second of these relations is crucial for current conservation.
Let us remark that on the light--cone these relations have to be 
written by using the interior derivative \cite{BT}, namely
\bea
\partial_\alpha \rightarrow d_\alpha 
\equiv 
(1+\xx \tilde\partial)\tilde\partial_\alpha 
- \hbox{\large$\frac{1}{2}$} \xx_\alpha \tilde{\partial}^2
\quad {\rm with} \quad d^2 = 0 \ .
\nonumber
\eea

Quite analogous relations hold for the pseudoscalar as well as
axial vector operators.
\section{MATRIX ELEMENTS OF TWIST--2 OPERATORS}
\renewcommand{\theequation}{\thesection.\arabic{equation}}
\setcounter{equation}{0}
\label{sec-4}

\vspace{1mm}
\noindent
Our aim is to obtain an expression for the non--forward Compton amplitude.
Up to now we considered the representation of the $T$--product of currents
containing twist--2 non--local (axial) vector operators.
The next step will be to perform matrix elements of that $T$--product which,
because of Eq.~(\ref{optw2}), can be traced back to matrix elements of the 
(pseudo) scalar operators (\ref{ortl}). Let us consider these matrix elements
 first. They decompose into two parts having  a
 Dirac and  a Pauli structure,
respectively:
\begin{eqnarray}
\label{scme}
\hspace{-.7cm}&&
e^2
\bigl
\langle p_2, S_2 \left|
O^{\rm tw2}\left({x}/{2}, - {x}/{2}\right)
\right |p_1, S_1 \bigr \rangle
\\
\hspace{-.7cm}&&
\quad
=
i\,\overline{u}(p_2,S_2)(\gamma x) u(p_1,S_1) \,\times
\nonumber\\
\hspace{-.7cm}&&
\qquad
\int\! Dz e^{-ixp(z)/2} f(z_1,z_2, p_i p_j x^2, p_i p_j,\mu^2_R) 
\nonumber\\
\hspace{-.7cm}&&
\quad
+\,i\,\overline{u}(p_2,S_2)(x \sigma p_-)  u(p_1,S_1) \,\times
\nonumber\\
\hspace{-.7cm}&&
\qquad
\int\! Dz e^{-ixp(z)/2}
 g(z_1,z_2 , p_i p_j x^2, p_i p_j,\mu^2_R)\ ,
\nonumber
\end{eqnarray}
where $(x \sigma p_-)\equiv x^\alpha \sigma_{\alpha\beta} p_-^\beta$
and $p(z) \equiv p_1z_1 + p_2z_2$, 
 $\mu_R$ denotes the renormalization scale and
\begin{eqnarray}
\label{Dz}
Dz
= \hbox{\large$\frac{1}{2}$} dz_1 dz_2
 \theta(1\!-\!z_1) \theta(1\!+\!z_1) \theta(1\!-\!z_2) \theta(1\!+\!z_2).
\nonumber
\end{eqnarray}
The kinematic decomposition given above follows if one takes into
account that the spinors $u(p_i,S_i) $ describing the hadrons satisfy
the free Dirac equation. The functions $
f(z_1,z_2, (p_i p_j )x^2, (p_i p_j),\mu^2_R) $ are the parton 
distribution
amplitudes and $z_i$ are the momentum fractions. For brevity we
drop the remaining variables. In the present approach we, moreover, set
$(p_i.p_j) \approx 0$. Under these assumptions the relation (\ref{co1}) 
is also valid for the matrix elements:
\begin{eqnarray}
\label{co1b}
\lefteqn{
\Box \langle p_2,S_2| O^{\rm tw2}( x/2,-x/2 )|p_1,S_1\rangle\ =
}\\
& \langle p_2,S_2|\Box O^{\rm tw2}( x/2,-x/2 )|p_1,S_1\rangle\ = 0\ .
\nonumber
\end{eqnarray}

Now, let us reconstruct the vector operator using Eq.~(\ref{optw2}). 
We obtain 
\begin{eqnarray}
\label{eqOM}
&&\hspace{-.7cm}
e^2
\langle p_2,S_2 \left|O^{\rm tw2}_\alpha
\left(x/2,-x/2\right)\right |p_1,S_1 \rangle  \\
&&\hspace{-.7cm} 
\quad
=\,i\int Dz\, e^{-i x p(z)/2} F(z_1,z_2)\,\times
\nonumber\\
&&\hspace{-.7cm}
\qquad\quad
[\overline{u}(p_2,S_2)\gamma_\alpha  u(p_1,S_1) 
\nonumber\\
&&\hspace{-.7cm}
\qquad\qquad
-\hbox{\large$\frac{i}{2}$} p_\alpha(z)  
\overline{u}(p_2,S_2)(\gamma  x)  u(p_1,S_1) ]
\nonumber\\
&&\hspace{-.7cm}
\quad
+\,i\int Dz\, e^{-i x p(z)/2} G(z_1,z_2) \, \times
\nonumber\\
&&\hspace{-.7cm}
\qquad\quad
[\overline{u}(p_2,S_2)\sigma_{\alpha\beta}{p_-^\beta} u(p_1,S_1)
\nonumber\\
&&\hspace{-.7cm}
\qquad\qquad
-\hbox{\large$\frac{i}{2}$}
 p_\alpha(z) \overline{u}(p_2,S_2)(x\sigma{p_-}) u(p_1,S_1)] \,,
\nonumber
\end{eqnarray}
where
\begin{eqnarray}
\label{fF}
F(z_1, z_2)=\int_0^1 \frac{d\lambda}{\lambda^2}\,
f\left(\frac{z_1}{\lambda},\frac{z_2}{\lambda}\right)\,,
\end{eqnarray}
and an analogous representation connects $G$ to $g$. Moreover,
similar representations between the corresponding functions 
$f_5$,\ $F_5$,\ $g_5$ and $G_5$ are valid for the operators containing 
$\gamma_5 $. Note that also here the necessary conditions
\begin{eqnarray}
\label{co1bv}
&&\hspace{-.7cm}
\Box \langle p_2,S_2| O^{\rm tw2}_{(5)\alpha}( x/2,-x/2 )|p_1,S_1 \rangle\,
= 0
\\
&&\hspace{-.7cm}
\label{co1vv}
\partial^\alpha \langle p_2,S_2| 
O^{\rm tw2}_{(5)\alpha}( x/2,-x/2 )|p_1,S_1\rangle\,
= 0
\end{eqnarray}
are satisfied. The relations (\ref{fF}) are of {\it central importance}
since they form
the theoretical basis of the Wandzura--Wilczek relations and allow
expressions for the non--forward Compton amplitude based on expectation
values of scalar operators only.
%
%
%
\section{CURRENT CONSERVATION}
\renewcommand{\theequation}{\thesection.\arabic{equation}}
\setcounter{equation}{0}
\label{sec-5}
\vspace{1mm}
\noindent
Current conservation is a very important criterion for the relevance of 
the derived expressions. Formally we have to start with Eq.~(\ref{COMP}), 
and apply Eq.~(\ref{Tmunur2}) and the expressions for the matrix elements
(\ref{eqOM}). In the case of forward scattering current conservation
holds~\cite{JBT}.

For non--forward scattering one is confronted with the following problem.
Let us 
consider the asymmetric current product with respect to $x=0$
\begin{eqnarray}
\hspace{-.7cm}
&& 
 \widehat{T}_{\mu\nu}^{\rm tw2}(\kappa_1 x, \kappa_2 x)  =
 - e^2 \frac{x^\lambda}{i \pi^2 (x^2-i\epsilon)^2}\,\times
\\
\hspace{-.7cm}
&& \quad
 \big[S^{\alpha}_{~ \mu\lambda \nu} 
 O^{\rm tw2}_\alpha(\kappa_1 x, \kappa_2 x )
-i \varepsilon^\alpha_{~\mu\lambda \nu} 
O_{5\ \alpha}^{\rm tw2}  (\kappa_1 x, \kappa_2 x)
\big],
\nonumber
\end{eqnarray}
where $\kappa_1 - \kappa_2 = 1$.
It is easy to convince oneself that, independent of the values of
$\kappa_i$, 
\bea
\label{J1}
\partial^\mu_x \widehat{T}_{\mu\nu}(\kappa_1 x, \kappa_2 x)=0=
\partial^\nu_x \widehat{T}_{\mu\nu}(\kappa_1 x, \kappa_2 x),
\eea
holds
because the relations (\ref{co3}) and (\ref{co2}) are satisfied.
These relations, ensuring tracelessness of the vector operators of
definite twist, are not changed by perturbation theory and 
renormalization.

This proves conservation of the first (or second) current if 
$\kappa_1 =1, \kappa_2 =0$ (or $\kappa_1 =0, \kappa_2 =-1$) is chosen. 
However, if conservation of both electromagnetic currents simultaneously
shall be proven  we have to study
\begin{eqnarray}
\hspace{-.7cm}
\label{Txy}
&&i RT[J_\mu(x) J_\nu (y) S]^{\rm tw2} =
 -e^2 \frac{\xi^\lambda}{i \pi^2 (\xi^2-i\epsilon)^2}\ \times
\nonumber\\
\hspace{-.7cm}
&& \qquad\qquad
[S^\alpha_{~\mu \lambda \nu}
O_{\alpha}^{\rm tw2}(\hbox{$\eta+\large\frac{\xi}{2}$},
\hbox{$\eta-\large \frac{\xi}{2}$}) 
\nonumber
\\
\hspace{-.7cm}
&&\qquad\qquad
- i \epsilon^\alpha_{~\mu \lambda \nu} 
 O_{5\ \alpha}^{\rm tw2}(\hbox{$\eta+\large\frac{\xi}{2}$},
\hbox{$\eta-\large \frac{\xi}{2}$})]\,,
\end{eqnarray}
where $\eta = (x+y)/2$ by convention denotes the reference point for 
the twist decomposition, and $\xi = x-y$ approaches the light--cone.
Because of
\bea
O_{(5)\alpha}^{\rm tw2}\big(\hbox{$\eta+\large\frac{\xi}{2}$},
\hbox{$\eta-\large \frac{\xi}{2}$}\big)
:=
e^{-i\eta P}\, O_{(5)\alpha}^{\rm tw2}\big(\hbox{$\large\frac{\xi}{2}$},
\hbox{$-\large \frac{\xi}{2}$}\big)\, e^{i\eta P}\,,
\nonumber
\eea
where $P_\mu$ is the momentum operator,
we find that Eqs.~(\ref{J1}) hold with respect to
the variable $\xi$ which is part of $x$ {\em and} $y$. 
Therefore, applying both derivations, either
$\frac{\partial}{\partial x^\mu} 
= \frac{1}{2} \frac{\partial}{\partial \eta^\mu} 
+ \frac{\partial}{\partial \xi^\mu}$
or
$\frac{\partial}{\partial y^\mu} 
= \frac{1}{2} \frac{\partial}{\partial \eta^\mu} 
- \frac{\partial}{\partial \xi^\mu}$, 
to the expression (\ref{Txy}) there remains, 
in both cases, a non--vanishing part which is proportional to $
[ P_\mu, O_{(5) \alpha}^{\rm tw2}(\hbox{$\eta +\large\frac{\xi}{2}$},
\hbox{$\eta-\large \frac{\xi}{2}$})]$.

This shows that the proof of current conservation essentially depends
on the choice of the reference point for the twist definition:  
translation of the non--linear operator also shifts that reference
point. It prevented us from proving conservation of both currents
simultaneously. This intrinsic problem of all the twist definitions
could be very important when non--leading twist
contributions are considered. 

Let us turn to the case of twist 2 now. For the explicit calculations of 
the resulting expressions we use a normalized helicity basis, 
cf.~\cite{BR}, with
$ \varepsilon_{0 \mu}^{(i)} = {q_{i\mu}}/{\sqrt{|q_i^2|}} $ for
$q_i^2 < 0 $ resp.
$ \varepsilon_{0 \mu}^{(i)} = {q_{i\mu}}/({\sqrt{2}|q_{0i}|}) $ for
$q_i^2 = 0 $.
We have shown in Ref.~\cite{BR}, that the current violating contributions
are  of  $O(\nu^{-1/2} \times {\rm OME})$ or higher order. 
These terms are of
higher twist and have to be dealt with the operator matrix elements
of the higher twist operators.~\footnote{C. Weiss has proven
recently~\cite{CW} using the representation of~\cite{BB} that the terms
$\propto O(\nu^{-1/2})$ cancel with corresponding terms due to
twist--3 operators.}
\section{INTEGRAL RELATIONS}
\renewcommand{\theequation}{\thesection.\arabic{equation}}
\setcounter{equation}{0}
\label{sec-6}

\vspace{1mm}
\noindent
Here we present the final expression for the Compton amplitude. The
calculations are given in \cite{BR}. We split the amplitude into its
symmetric part and the antisymmetric part 
$T^{\mu  \nu} = T^{\mu \nu}_{\rm s} + T^{\mu \nu}_{\rm as} $. 
In the case of forward
scattering the former one corresponds to the unpolarized and the latter
one to the polarized contribution.
First we consider the symmetric part:
%
\begin{eqnarray}
\label{eqAmn}
&&\hspace{-.7cm}
T^{\mu\nu,{\rm tw2}}_{\rm s} 
= - \frac{1}{\nu} \int_{-1}^1\!\!
     dt \frac{1}{\xi+t -i\varepsilon }\,\times \\
&&\hspace{-.7cm}\quad
 \left\{\Big[2
\left(g^{\mu\nu}(qp_+) - (q^\mu p_+^\nu + q^\nu p_+^\mu) \right)
 f_1(t,\eta)\right.
\nonumber \\ 
&&\hspace{-.7cm}\qquad
+\,  p_+^\mu p_+^\nu f_2(t,\eta)\Big] \overline u(p_2,S_2)
 (\gamma q) u(p_1,S_1) 
\nonumber \\ 
&&\hspace{-.7cm}\quad
+\Big[2\left(g^{\mu\nu}({qp_+}) - ({q^\mu p_+^\nu + q^\nu p_+^\mu })\right)
 g_1(t,\eta)
\nonumber \\ 
&&\hspace{-.7cm}\qquad
+ \left. {p_+^\mu p_+^\nu} g_2(t,\eta)\Big] \overline u(p_2,S_2)
 (q \sigma p_-) u(p_1,S_1) \right\}
\nonumber \\
&&\hspace{-.7cm}\quad 
+\, {\rm non-leading~terms}~,
\nonumber
\end{eqnarray}
where $ t = z_+ +\eta z_-, \, z_\pm= \frac{1}{2}(z_2 \pm z_1) $.
Here the partition functions $f_i$ and $g_i$ are `one--variable' 
distribution amplitudes which are defined by 
\begin{eqnarray}
f_{(5)}(t,\eta)\!\!\!&=&
\!\!\!\!\int dz_-\,f_{(5)}(z_+ = t -\eta z_-, z_-),\\
g_{(5)}(t,\eta)\!\!\!&=&
\!\!\!\!\int dz_-\,g_{(5)}(z_+ = t -\eta z_-, z_-),
\end{eqnarray}
from the `two--variable' distribution amplitudes used in the 
representation (\ref{scme}) of the matrix elements of the (pseudo) 
scalar operators. Unlike the case of forward scattering these functions
do not depend on scaling variables only but besides of the scaling
variable $\eta$ which describes non--forwardness of the combination
of momentum fractions $t$. The following new relations are obtained 
between the amplitude--functions $f_i (g_i)$, see~\cite{BR}~:
\begin{eqnarray}
   f_2(t,\eta)\!\!&=&\!\!2tf_1(t,\eta)\equiv 2 f(t,\eta), \\
   g_2(t,\eta)\!\!&=&\!\!2tg_1(t,\eta)\equiv 2 g(t,\eta)~.
\end{eqnarray}
These relations are structurally similar to the {\sc Callan--Gross}
relation for forward scattering. There, by virtue of the optical
theorem,
$ \frac{1}{\xi + t + i\epsilon} \rightarrow i \pi \delta(t + \xi ) $
and for $p_2 \rightarrow p_1 =p $ it follows $ t  \rightarrow z_+ \equiv
z$, $t$ is turned into a scaling variable. In the above expressions
furthermore
\bea 
\overline u(p_2,S_2)  (\gamma q) u(p_1,S_1) &\rightarrow& 2pq, 
\nonumber\\
\overline u(p_2,S_2) (q \sigma p_-) u(p_1,S_1)& \rightarrow& 0
\nonumber
\eea 
holds in the latter case.

The result for the antisymmetric part is more complicated:
\begin{eqnarray}
T_{(\lambda_1)(\lambda_2)}
  &=&  \epsilon_{\mu (\lambda_1)}^{(2)}
     \epsilon_{\nu (\lambda_2)}^{(1)}
     T^{\mu\nu,{\rm tw2}}_{\rm as} 
\nonumber\\
  &=&  i \epsilon^{\mu \rho\nu\sigma}
     \epsilon_{\mu (\lambda_1)}^{(2)}
     \epsilon_{\nu (\lambda_2)}^{(1)} 
     B_{\rho \sigma}
\end{eqnarray}
with
\begin{eqnarray}
&&\hspace{-.7cm}
B^{\rho\sigma} = -  \frac{q^\rho}{\nu^2}\Big[
\int_{-1}^1 dt \frac{1}{\xi+t -i\varepsilon}\,\times 
\nonumber\\
&&\hspace{-.7cm} \quad
 \Big\{ \Big(f_{5,1}(t,\eta) + f_{5,2}(t,\eta)\Big){\nu} S^{12}_\sigma
 \nonumber \\
&&\hspace{-.7cm} \quad
+ \Big(g_{5,1}(t,\eta) + g_{5,2}(t,\eta)\Big){\nu} \Sigma^{12}_\sigma
  \nonumber \\
&&\hspace{-.7cm} \quad
+f_{5,2}(t,\eta) p_{+}^{\sigma}(q S^{12})
+g_{5,2}(t,\eta) p_{+}^{\sigma}(q \Sigma^{12})
  \Big\} \nonumber
 \Big] .
\nonumber
\end{eqnarray}
Here, we used the following abbreviations for the spinor structure
\begin{eqnarray}
 S^{12}_\sigma
        \!\!\! &=& \!\!\! 
-\hbox{\large$\frac{1}{2}$}  
\overline u(p_2,S_2) \gamma_5 \gamma_\sigma  u(p_1,S_1) , 
\nonumber\\
 \Sigma^{12}_\sigma
           \!\!\!  &=& \!\!\! -
\hbox{\large$\frac{1}{2}$}  
\overline u(p_2,S_2)\gamma_5 \sigma_{\sigma \rho }p_-^\rho u(p_1,S_1).
\nonumber
\end{eqnarray}
In the case of forward scattering $p_2 \rightarrow p_1= p$,
$ S^{12}_\sigma \rightarrow S_\sigma $       and
$ \Sigma^{12}_\sigma \rightarrow 0 $ holds,
where $S_\sigma$ is the spin vector
introduced for forward scattering. We obtain the following relations
for the amplitude functions, cf.~\cite{BR}~:
\begin{eqnarray}
   f_{5,1}(t,\eta) \!\!\! &\equiv& \!\!\! f_5(t,\eta), 
\\
   f_{5,2}(t,\eta) \!\!\! &=& \!\!\! -f_{5}(t,\eta)
     +\! \int_t^{{\rm sgn}\ t}\!\!\!\! dz\,
 \frac{f_{5}(z,\eta)}{z} , 
\\
   g_{5,1}(t,\eta) \!\!\! &\equiv& \!\!\! g_5(t,\eta), 
\\
   g_{5,2}(t,\eta) \!\!\! &=& \!\!\! -g_{5}(t,\eta)
     +\! \int_t^{{\rm sgn}\ t}\!\!\!\! dz \,
 \frac{g_{5}(z,\eta)}{z}~.
\end{eqnarray}
Again $f_{5,i}$ and $g_{5,i}$ depend on the momentum fraction $t$
and a scaling variable. These relations generalize the 
{\sc Wandzura--Wilczek} relation of deep inelastic scattering to 
non--forward amplitudes.
\section{CONCLUSIONS}
\renewcommand{\theequation}{\thesection.\arabic{equation}}
\setcounter{equation}{0}
\label{sec-7}
We studied the structure of the virtual Compton amplitude for
deep--inelastic non--forward scattering $\gamma^* + p \rightarrow
\gamma^{\prime *} +p'$ in lowest order in QED in the massless limit. 
In the
generalized Bjorken region $(qp_+),\, -q^2 \rightarrow \infty$ the
twist--2 contributions to the Compton amplitude were calculated using
the non--local operator product expansion. The twist separation and
the relations between twist--2 vector operators and twist 2 scalar
operators are essential for the current conservation at the level of
twist--2 and, moreover, all parton distributions are connected with the
matrix elements of the scalar twist--2   operators.

The relations between the twist--2 contributions of the unpolarized and 
polarized amplitude functions were derived. They are the non--forward 
generalizations of the {\sc Callan}--{\sc Gross} and 
{\sc Wandzura}--{\sc Wilczek} relations for unpolarized and polarized 
deep--inelastic forward scattering. The relations for the {\sc Dirac} and
{\sc Pauli} parts are of the same form.

\vspace{2mm}
\noindent
{\bf Acknowledgement.} This work was supported in part by EU contract
FMRX-CT98-0194 (DG 12--MIHT). B.G. and D.R. would like to thank DESY
for financial support.


\begin{thebibliography}{999}
%
\bibitem{BGR}
J. Bl\"umlein, B. Geyer, and D. Robaschik, Nucl. Phys. {\bf B560} (1999)
283.
%
\bibitem{BR}
J. Bl\"umlein and D. Robaschik, {\tt hep-ph/ 0002071}, 
Nucl. Phys. {\bf B} in print.
%
\bibitem{LEIP}
F.~Dittes, B.~Geyer, J.~Ho\v{r}ej\v{s}i, D. M\"uller, and D.~Robaschik,
Fortschr. Phys.  {\bf 42} (1994) 2.
%
\bibitem{ADI}
A.V.~Radyushkin, Phys. Lett. {\bf B385} (1996) 333; Phys. Rev. {\bf D56} 
(1997) 5524;\\
X.~Ji, Phys. Rev. Lett. {\bf 78} (1997) 610; Phys. Rev. {\bf D55} (1997)
7114;\\
L. Mankiewicz, G. Piller, and T. Weigl, Eur. J. Phys. {\bf C5} (1998) 
119.
%
\bibitem{GLR}
B. Geyer, M.~Lazar, and D. Robaschik, Nucl. Phys. {\bf B559} (1999) 339.
%
\bibitem{GL}
B. Geyer and M.~Lazar, {\tt hep-th/0003080}.
%
\bibitem{NLC}
S.A. Anikin and O.I. Zavialov, Ann. Phys. (NY) {\bf 116} (1978)
135;\\
R.A.~Brandt and G.~Preparata, Fortschr. Phys. {\bf 18} (1970)
249.
%
\bibitem{TW}
D.J.~Gross and S.B.~Treiman, Phys. Rev. {\bf D4} (1971) 1059.
%
\bibitem{RC}
D.J. Gross and F. Wilczek, Phys. Rev. {\bf D8} (1973) 3633; {\bf D9}
(1974) 980;\\
H. Georgi and D. Politzer, Phys. Rev. {\bf D9} (1974) 416.
%
\bibitem{BB}
I.I.~Balitsky, Phys. Letters {\bf 124B} (1983) 230,\\
I.I.~Balitsky and V.M.~Braun, Nucl. Phys. {\bf B311} (1988/89) 541.
%
\bibitem{BT}
V. Bargmann and I.T. Todorov, J. Math. Phys. {\bf 18} (1977) 1141.
%
\bibitem{JBT}
J. Bl\"umlein and A. Tkabladze, Nucl. Phys. {\bf B553} (1999) 427.
%
\bibitem{CW}
C. Weiss, Bochum Note, February 2000.
%
\bibitem{CG}
C.G. Callan and D.J. Gross, Phys. Rev. Lett. {\bf 22} (1969) 156.
%
\bibitem{WW}
S. Wandzura and F. Wilczek, Phys. Lett. {\bf B72} (1977) 95.
\end{thebibliography}
\end{document}